\documentclass[useAMS,usenatbib,usegraphicx]{mn2e}
\usepackage{times}
\input{psfig.sty}

\def\gs{\mathrel{\raise0.35ex\hbox{$\scriptstyle >$}\kern-0.6em
\lower0.40ex\hbox{{$\scriptstyle \sim$}}}}
\def\ls{\mathrel{\raise0.01935ex\hbox{$\scriptstyle <$}\kern-0.6em
\lower0.40ex\hbox{{$\scriptstyle \sim$}}}}

\pagerange{\pageref{firstpage}--\pageref{lastpage}} \pubyear{2005}

\title[Nature of extremely bright SDSS LBGs]
{Rest-frame optical and far-infrared observations of extremely bright
Lyman-break galaxy candidates at {\bf\em z}\,$\sim$\,2.5}

\author[Ivison et al.]{R.\,J.\ Ivison,$^{\! 1,2}$
Ian Smail,$^{\! 3}$
Misty Bentz,$^{\! 4}$
J.\,A.\ Stevens,$^{\! 1,5}$
K.\ Men\'{e}ndez-Delmestre,$^{\! 6}$ \and
S.\,C.\ Chapman$^{6}$
and
A.\,W.\ Blain$^{6}$\\
$^1$ UK Astronomy Technology Centre, Royal Observatory,
Blackford Hill, Edinburgh EH9 3HJ\\
$^2$ Institute for Astronomy, University of Edinburgh,
Blackford Hill, Edinburgh EH9 3HJ\\
$^3$ Institute for Computational Cosmology, University of Durham,
South Road, Durham DH1 3LE\\
$^4$ Department of Astronomy, The Ohio State University,
140 W.\ 18th Ave, Columbus, OH 43210-1173, USA\\
$^5$ Centre for Astrophysics Research, Science and Technology Research Centre,
University of Hertfordshire, College Lane, Hatfield AL10 9AB\\
$^6$ Department of Astronomy, California Institute of Technology,
MS 320-47, Pasadena, CA 91125, USA}

\begin{document}

\maketitle

\setcounter{footnote}{5}

\begin{abstract}
We have investigated the rest-frame optical and far-infrared
properties of a sample of extremely bright candidate Lyman-break
galaxies (LBG) identified in the Sloan Digital Sky Survey. Their high
ultraviolet luminosities and lack of strong ultraviolet emission lines
are suggestive of massive starbursts, although it is possible that
they are more typical luminosity LBGs which have been highly magnified
by strong gravitational lensing. Alternatively, they may be an unusual
class of weak-lined quasars. If the ultraviolet and submillimetre
properties of these objects mirror those of less luminous, starburst
LBGs, then they should have detectable rest-frame far-infrared
emission. However, our submm photometry fails to detect such emission,
indicating that these systems are not merely scaled-up (either
intrinsically or as a result of lensing) examples of typical LBGs. In
addition we have searched for the morphological signatures of strong
lensing, using high-resolution, near-infrared imaging, but we find
none. Instead, near-infrared spectroscopy reveals that these systems
are, in fact, a rare class of broad absorption-line (BAL) quasars.
\end{abstract}


\section{Introduction}

The very large areal coverage of the Sloan Digital Sky Survey
\citep[SDSS,][]{york00} provides a unique opportunity to identify rare,
intrinsically luminous examples of high-redshift galaxy populations, as well as
similarly rare, strongly-gravitationally magnified examples of more typical
high-redshift galaxies.  To exploit this opportunity, \citet{bo04}
searched the SDSS Early Data Release \citep{sto02} for unusual quasars with
anomalously low C\,{\sc iv}\,154.9nm emission, and found a luminous
$z\sim\rm 2.5$ starburst candidate that appeared to have been incorrectly
classified as a quasar by the SDSS pipeline.

Following on from this find, \citet{b04} identified a further five sources from
the SDSS First Data Release (DR1) Quasar Catalog \citep{sch03} at
$z$\,$\sim$\,2.5--2.8 with rest-frame ultraviolet (UV) colours similar to
Lyman-break galaxies \citep[LBGs,][]{s03} and exhibiting weak or absent
high-ionisation emission lines in their rest-frame UV spectra. All six objects
have $r$-band magnitudes of 19.8--20.5, with a median of $r\sim\rm 20.3$, i.e.\
they are an order of magnitude brighter than the most luminous objects in
existing LBG surveys. \citet{b04} showed that if their UV emission arises
solely from star formation then they are intensely active star-forming galaxies
with strong lower limits on their star-formation rates ranging from 300 to
1100\,M$_{\odot}$\,yr$^{-1}$, assuming negligible absorption by dust and
adopting a continuous star-formation rate over $\rm 10^8$\,yrs \citep{ken98}.
Thus these candidate LBGs could be rare, extreme starbursts seen at the epoch,
$z\sim\rm 2.5$, when both the accretion luminosity density and the
star-formation rate density in the Universe are believed to peak
\citep{m00,cha05}.

Alternatively, the brightness and apparent rarity of these systems could simply
reflect the fact that they are rare, strongly-magnified examples of the
normal-luminosity LBG population. The LBG candidates show no evidence of
multiple components at the resolution of SDSS imaging ($\gs$1\,arcsec) and
there are no obvious foreground lensing structures. Nevertheless, gravitational
lensing cannot yet be ruled out based on existing imaging.

The best argument {\it against} these galaxies being highly-magnified
examples of normal LBGs results from their spectral properties: the
underlying continuua are much redder than typical LBGs, with observed
200--600-nm spectral indices ranging from $\alpha=\rm -1.9$ to $\rm
-2.5$ (where $F_\nu \propto \nu^{+\alpha}$), cf.\ $\alpha\rm >-1.7$
for most LBGs \citep{shap03}, and the low-resolution SDSS spectra show
strong (sometimes broad) interstellar absorption lines and Ly$\alpha$
emission in several cases, as well as hints of broad C\,{\sc
iii}]\,190.9nm emission in two cases \citep{b04}.  These features
contrast with the narrow emission and interstellar absorption lines
seen in the composite LBG spectrum of \citet{shap03}, but may be
explained due to star-formation activity (and perhaps associated
winds) an order of magnitude more vigorous than the sample considered
by \citet{shap03}.  The gross spectral properties of a source should
not be affected markedly by lensing, so these differences argue that
these galaxies are unlikely to be highly-magnified examples of the
general LBG population. Rather, the spectral properties of these
sources share key characteristics with the submm-selected galaxies
identified in the recent spectroscopic survey of
\citet{cha03,cha05}. In particular, they resemble N2\,850.4, a
composite starburst/AGN at $z$\,=\,2.38 \citep{smail03} and the
BAL--Sy2/QSO SMM J02399$-$0136 at $z$\,=\,2.80 \citep{ivi98, vc01} ---
the two-component Ly$\alpha$ emission, broad C\,{\sc iii}] and the
absorption seen in C\,{\sc ii}, Si\,{\sc iv}, Al\,{\sc iii} and
C\,{\sc iv} --- although the candidate LBGs lack such prominent
P-Cygni profiles. The LBG candidates also share some common
characteristics with broad absorption-line (BAL) quasars --- the
presence of broad C\,{\sc iii}]190.9nm in two examples and some very
broad absorption lines --- although the line profiles are not typical
of BALs.  There is thus a possibility that the UV emission from these
galaxies is powered by accretion rather than star formation, or that
the sample is a heterogeneous mix of star-forming galaxies and active
galactic nuclei (AGN).  A recent paper by \citet{hall04} discusses an
unusual object, SDSS\,J113658.36+024220.1, whose rest-frame UV
spectrum shows a single emission line corresponding to Ly$\alpha$ but
no obvious metal-line emission, which bears some similarities to the
candidate LBGs studied here. \citet{hall04} interpret SDSS\,J1136 as
an AGN based in large part on tentative optical variability and its
strong radio emission, $\sim$1.4\,mJy at 1.4\,GHz.

If these galaxies are truly related to starburst LBGs then their prodigious
star formation should be betrayed in the rest-frame far-infrared.  Here, we
exploit Submm Common-User Bolometer Array --- SCUBA, \citet{hol99} ---
submillimetre (submm) photometry to search for such emission.  We then use new,
high-resolution, near-infrared imaging to identify the morphological signatures
of strong lensing. Finally, we present near-infrared spectra of these galaxies,
covering a number of key rest-frame optical emission lines falling in the $H$
and $K$ atmospheric windows, to spectroscopically classify the galaxies and to
search for quasar signatures such as a tell-tale broad component to the
H$\alpha$ line.  We describe our observations in \S2, present our results and
discussion in \S3 and give our conclusions in \S4.

%
%
\begin{figure*}
\centerline{\psfig{file=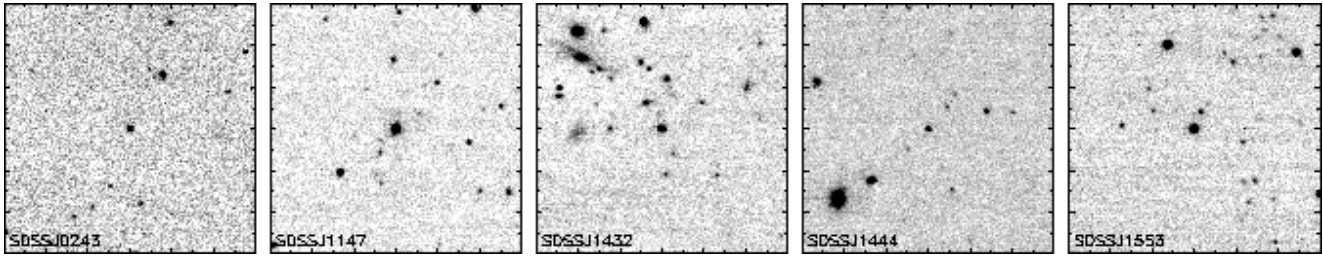,angle=270,width=17.5cm}}
\caption{Grey-scale representations of our deep $K$-band imaging of the five
luminous SDSS LBG candidates. Note the cluster of faint $K$-band sources around
SDSS\,J1147 and the foreground group of bright galaxies near SDSS\,J1432.  Each
image is 60\,$\times$\,60\,arcsec with North to the top and East to the left and has
been smoothed by a Gaussian with a FWHM of 0.2\,arcsec. }
\label{fig1}
\end{figure*}

%
%
\begin{figure*}
\centerline{\psfig{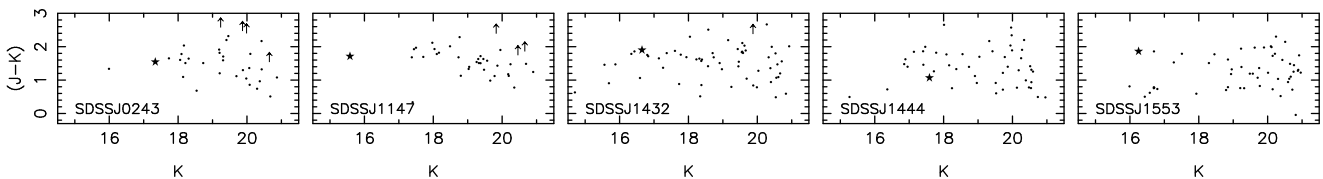}}
\caption{Distribution of sources on the $(J-K)$--$K$ colour-magnitude plane for
each of our five target fields.  We identify the candidate LBG in each field by
an $\ast$.  We see a strong sequence in the distribution of galaxy colours in
the field of SDSS\,J1432 with a typical colour of $(J-K)\sim\rm 1.8$, suggesting
that the over-density of galaxies in this field is at $z\sim\rm 0.8$. We plot
only those galaxies with total $K$-band magnitudes brighter than the 5-$\sigma$
limit of $K=\rm 21.0$ and show lower limits for those sources with
$<$3-$\sigma$ detections in our 2\,arcsec-diameter photometric aperture in the
$J$-band.  The total area surveyed across the five fields is 13.3\,arcmin$^2$.}
\label{fig2}
\end{figure*} 

\section{Submillimetre and near-infrared observations}

%
%
\begin{table*}
\begin{center}
\caption{Near-infrared, submm and radio properties of the SDSS LBG candidates}
\begin{tabular}{lccccccc}
\hline
Source name & $z$(UV)$^a$&$K^b$ & $(J-K)^b$ & $(i-K)$ & $(r-K)$ & $S_{850\mu \rm
m}$&$S_{1.4\rm GHz}$\\
&& (mag) & (mag) & (mag) & (mag) & (mJy) & (mJy)\\
SDSS\,J024343.77$-$082109.9&2.590
& 17.33\,$\pm$\,0.02&1.62\,$\pm$\,0.02& 2.75    & 3.07    &$-$2.13\,$\pm$\,2.02&5$\sigma$\,$<$\,0.97\\
SDSS\,J114756.00$-$025023.5&2.556
&15.58\,$\pm$\,0.02&1.71\,$\pm$\,0.03& 4.25 & 3.71 & +2.61\,$\pm$\,2.90&5$\sigma$\,$<$\,1.03\\
SDSS\,J134026.44+634433.2$^c$&2.786
& 16.90\,$\pm$\,0.10 & 1.00\,$\pm$\,0.15 & 2.06     & 2.76 &+1.93\,$\pm$\,3.33&5$\sigma$\,$<$\,0.98\\
SDSS\,J143223.10$-$000116.4&2.472
&16.65\,$\pm$\,0.03&1.90\,$\pm$\,0.04& 3.86 & 3.47 & ... &5$\sigma$\,$<$\,0.97\\
SDSS\,J144424.55+013457.0&2.670
&17.59\,$\pm$\,0.04&1.07\,$\pm$\,0.05& 2.92 & 2.49 & $-$1.33\,$\pm$\,2.78&5$\sigma$\,$<$\,1.01\\
SDSS\,J155359.96+005641.3&2.635
&16.25\,$\pm$\,0.03&1.86\,$\pm$\,0.04& 3.95 & 3.43&  ... &5$\sigma$\,$<$\,0.94\\
\hline
\end{tabular}

{\small
$^{a}$ UV redshifts were measured using cross-correlation with a
quasar template and searches for emission lines \citep{sch02}.\hfil \\
$^{b}$ Aperture magnitudes, measured in 6-arcsec-diameter apertures.
Correction for line contamination is discussed in \S3.2.\hfil \\
$^{c}$ Near-infrared photometry from \citet{tep04}.\hfil\\ 
}
\end{center}
\end{table*}

Submm photometry observations were obtained for four of our LBG candidates in
service time on 2004 January 15 and 28 with SCUBA on the James
Clerk Maxwell Telescope (JCMT\footnote{The JCMT is operated by the Joint
Astronomy Centre in Hilo, Hawaii, on behalf of the Particle Physics and
Astronomy Research Council (PPARC) in the UK, the National Research
Council of Canada, and The Netherlands Organisation for Scientific
Research.}). On the first night, observations of SDSS\,J1147, SDSS\,J1340 and
SDSS\,J1444 were made in average opacity conditions ($\tau_{850\mu\rm
m}$\,$\sim$\,0.3--0.4), and on the second night observations of SDSS\,J0243
were made in better conditions ($\tau_{850\mu\rm m}$\,$\sim$\,0.1--0.2).  We
used SCUBA in two-bolometer mode, giving a $\sim$15 per cent improvement in
signal-to-noise compared with one-bolometer mode. Each source was observed for
1.8\,ks. The {\sc starlink} package {\sc surf} was used to reduce the data for
each bolometer separately. The resulting signals were then calibrated against
the JCMT secondary calibrators CRL\,618 and 16293$-$2422, and co-added to give
weighted 850-$\mu$m flux densities and errors (see Table~1).  Calibration
uncertainties are estimated to be $\sim$10 per cent.

Near-infrared imaging data in the $J$- and $K$-bands were obtained for the five
bright LBG candidates accessible to the 3.8-m UK Infrared Telescope
(UKIRT\footnote{UKIRT is operated by the Joint Astronomy Centre on behalf of
PPARC.})  during 2004 January--April and 2004 July--August. Flexible scheduling
enabled us to utilise better-than-average seeing on Mauna Kea, 0.4--0.6\,arcsec,
and we employed the UKIRT Fast Track Imager, UFTI \citep{roc03}, a 1024$^2$
HgCdTe array with 0.091-arcsec pixels, to exploit those conditions. The total
integration time in each filter, built up whilst dithering every 60\,s, was
3.8\,ks. Contiguous observations of nearby faint standards were used to
determine zero points. The 3-$\sigma$ detection threshold is $K$\,$\sim$\,21.5
in a 4-arcsec-diameter aperture. Data were reduced using {\sc orac-dr} and we report
effective total magnitudes/colours (measured from 6-arcsec-diameter photometry)
for the LBG candidates in Table~1. Objects in the frames were then identified
and catalogued using SExtractor \citep{sex} and colours measured in
2-arcsec-diameter apertures from the aligned $J$- and $K$-band frames. $K$-band
images of regions around each target are shown in Fig.~1, with the $(J-K)$--$K$
colour-magnitude distributions for each field displayed in Fig.~2.

Spectra were obtained during 2004 April and August with the UKIRT 1--5\,$\mu$m
Imager Spectrometer, UIST \citep{ram98}, which utilises a 1024$^2$ InSb array
with 0.12-arcsec pixels. UIST's $HK$ grism was used to cover the
1.4--2.5\,$\mu$m region, with measured resolutions of $\lambda/\Delta \lambda$
= 390 and 680 (for arc lines at 1.5 and 2.3\,$\mu$m) for our 4-pixel-wide slit
(0.48\,$\times$\,120\,arcsec). Acquisition was accomplished using 20--60-s
sky-subtracted images of each field. We are confident that all targets were
placed within a pixel of the optimal position on the slit. Each target was
observed for 6.7\,ks, nodding along the slit in an A-B-B-A sequence every
240\,s. An Argon arc spectrum and a flatfield frame were obtained prior to
observations of each target. Nearby F5V standard stars were observed
contiguously to set the flux scale, having interpolated across their Hydrogen
absorption lines before ratioing. The frames were reduced using {\sc orac-dr},
with optimal extraction of spectra accomplished using {\sc figaro}.  For
comparison purposes we also obtained UIST $HK$-spectra of the weak-lined quasar
SDSS\,J113658.36+024220.1 \citep[$z$\,=\,2.4917,][]{hall04} and the
$z$\,=\,2.320 LoBAL quasar, SDSS\,J135317.80$-$000501.3 \citep{rei03,wil03}.
All the spectra are shown in Fig.~3.

A further spectrum was obtained for the LBG candidate inaccessible to UKIRT
(SDSS\,J134026.44+634433.2) using NIRSPEC on Keck-{\sc ii}\footnote{The W.\ M.\
Keck Observatory is operated as a scientific partnership among the California
Institute of Technology, the University of California and the National
Aeronautics and Space Administration. The Observatory was made possible by the
generous financial support of the W.\ M.\ Keck Foundation.} during relatively
poor conditions (0.9-arcsec seeing) on 2005 March 18. The full spectral range
was 2.27--2.69\,$\mu$m with a resolution of $\sim$1,500, utilising a
42\,$\times$\,0.76-arcsec (4-pixel) slit. The total integration time was
1.6\,ks, split into four A-B-B-A sequences with 100\,s per exposure. A flux
standard was not observed; otherwise, the data reduction followed that employed
for UIST.

\section{Results and discussion}

We now discuss the insights provided into the nature of the candidate LBGs by
our various multi-wavelength observations. The basic properties of our sample
--- redshifts, near-infrared photometry and 850-$\mu$m flux densities --- are
listed in Table~1. $K$-band imaging of the target fields are shown in Fig.~1,
$(J-K)$--$K$ colour magnitude diagrams of these fields in Fig.~2 and the $HK$
spectra of the candidate LBGs in Fig.~3. Line widths and fluxes for the
stronger features in these spectra are listed in Table~2.

\subsection{Submillimetre properties}

Studies of submm-selected galaxies (SMGs) have concluded that most are too
faint in the UV to be identified by the photometric selection used for
$z\sim\rm 3$ LBG surveys \citep{webb03}.  This suggests little overlap between
these two well-studied classes of high-redshift star-forming galaxies.
However, statistical measurements of the submm emission from samples of
$z\sim\rm 3$ LBGs, reaching below SCUBA's confusion limit, suggest that they
may contribute substantially to the SMG population at sub-mJy levels
\citep{pea00,webb03,kneib04}. Only a handful of brighter examples are known
\citep{cha02}. There is evidence of a more significant overlap between SMGs and
the UV-selected population identified at somewhat lower redshifts,
$z$\,$\sim$\,1.5--2.5 \citep{s04}, although again many of the SMGs are too
faint to be included in the photometric samples \citep{cha05}.

The raw star-formation rates estimated from the observed UV luminosities of the
SDSS sources, uncorrected for dust extinction, would imply $L_{\rm FIR}\gs\rm
10^{13}$\,L$_{\odot}$ for our sample, where SFR = $\epsilon$ 10$^{-10} L_{\rm
FIR}$\,M$_{\odot}$\,yr$^{-1}$ and $\epsilon$ = 0.8--2.1 \citep{sy83, tt86},
and thus 850-$\mu$m flux densities of 3--10\,mJy \citep{bl86}. Assuming
a correction factor for dust extinction typical of LBGs, 6$\times$
\citep{pet02,erb03}, the predicted submm fluxes would increase by a similar
factor. This suggests that the candidate LBGs should be detectable in the submm
waveband if they have submm/UV flux ratios similar to the more typical
luminosity members of this population. This conclusion holds whether these
galaxies are either intrinsically bright in the UV or are lensed (assuming that
the lensing does not preferentially boost the UV-bright regions).

The submm data presented in Table~1 demonstrate that the four SDSS sources we
have observed are all undetected individually at flux limits of 6--8\,mJy.
This implies that it is unlikely that these galaxies are simply scaled-up or
strongly-lensed examples of typical-luminosity $z\sim\rm 3$ LBGs.  Indeed, the
weighted mean for the four sources ($-$0.36\,$\pm$\,1.31\,mJy) suggests that
they would not have been detected in even the deepest submm survey, although we
cannot rule out the possibility that the sample is heterogeneous, with a
handful of faint submm emitters.  

For completeness we note that a search of the FIRST radio survey \citep{bwh95}
yielded only upper limits at 1.4\,GHz (Table~1), and that none of the galaxies
were detected by {\it ROSAT} to limits appropriate for the X-Ray All-Sky
Survey.

\subsection{Photometric and morphological properties}

%
%
\begin{table*}
\begin{center}
\caption{Spectral properties of the SDSS LBG candidates and comparison quasars}
\begin{tabular}{lccccccc}
\hline
Source & FWHM(H$\alpha$) & Flux(H$\alpha$) &  Flux([O{\sc iii}]5007) & Flux(H$\beta$) & EW$^a$ (H$\alpha$) & $z$([O{\sc iii}]) & $z$(H$\alpha$)\\
Name      & (km\,s$^{-1}$) & ($10^{-16}$\,W\,m$^{-2}$) & ($10^{-16}$\,W\,m$^{-2}$) & ($10^{-16}$\,W\,m$^{-2}$) & (nm) &&\cr 
SDSS\,J024343.77$-$082109.9
&5,360&1.4\,$\pm$\,0.1&3$\sigma$\,$<$\,0.3&3$\sigma$\,$<$\,0.3
&$-$96&2.5940&2.5995\\
SDSS\,J114756.00$-$025023.5
&13,830&5.3\,$\pm$\,0.2&0.5\,$\pm$\,0.1&0.4\,$\pm$\,0.1
&$-$152&2.5701&2.5668\\
SDSS\,J134026.44+634433.2
&12,200$^b$&...&...&...&$-$164$^b$&...&...\\
SDSS\,J143223.10$-$000116.4
&6,150&8.6\,$\pm$\,0.2&1.3\,$\pm$\,0.1&2.1\,$\pm$\,0.1
&$-$166&2.4728&2.4772\\
SDSS\,J144424.55+013457.0  
&5,730$^c$&2.1\,$\pm$\,0.3$^c$&0.6\,$\pm$\,0.1&3$\sigma$\,$<$\,0.4
&$-$30$^c$&2.6767&2.6715\\
SDSS\,J155359.96+005641.3  
&7,190&11.6\,$\pm$\,0.3&0.5\,$\pm$\,0.1&2.0\,$\pm$\,0.2
&$-$157&2.6350&2.6404\\
\\
SDSS\,J113658.36+024220.1
&4,100&3.5\,$\pm$\,0.1&0.3\,$\pm$\,0.1&0.4\,$\pm$\,0.1
&$-$106&2.4928&2.4946\\
SDSS\,J135317.80$-$000501.3
&5,180&6.1\,$\pm$\,0.2&3$\sigma$\,$<$\,6.0&3$\sigma$\,$<$\,0.3
&$-$139&--&2.3182\\
\hline
\end{tabular}

{\small
$^{a}$ Not corrected to the rest frame. \hfil \\
$^{b}$ Fits assume $z$(H$\alpha$) = $z$(UV) = 2.786. \hfil \\ 
$^{c}$ Lower limits, given the lack of continuum redward of the line. \hfil \\

}
\end{center}
\end{table*}

%
%
\begin{figure*}
\psfig{file=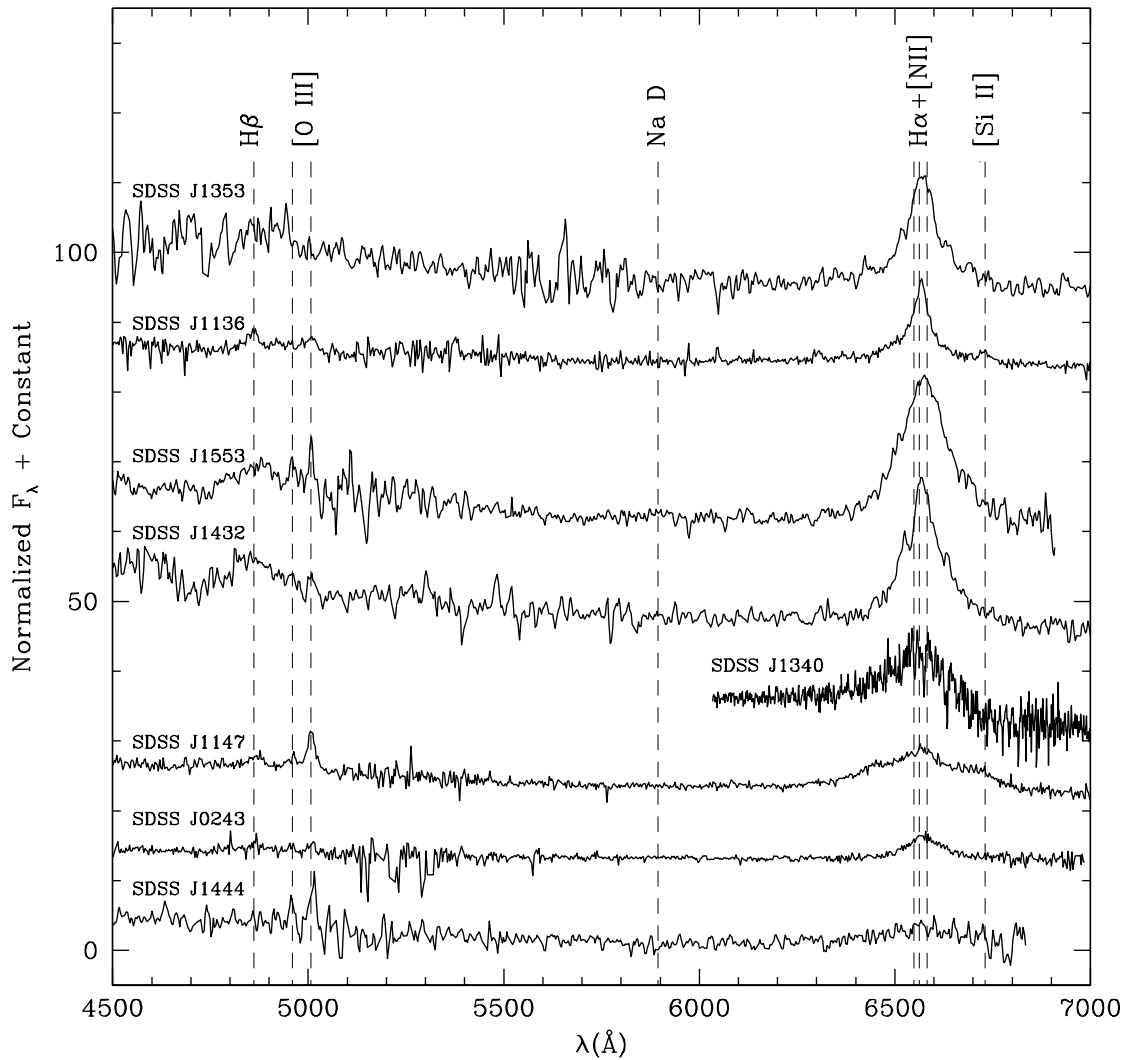,width=15.5cm,angle=0}
\caption{The six spectra at the bottom of this panel illustrate the rest-frame
optical spectra of the six luminous SDSS LBG candidates (arbitrarily offset in
flux). SDSS\,J1340, observed with NIRSPEC, has only $K$-band coverage.  Above
these, for comparison, are similar spectra of the LoBAL QSO, SDSS\,J1353, and
the Ly$\alpha$-only AGN, SDSS\,J1136. All the spectra are smoothed with a
Gaussian of $\sigma=\rm 0.75$\,pixels. We mark the wavelengths of possible
absorption or emission lines which may be visible in these galaxies.}
\label{fig3}

\end{figure*}

Our near-infrared observations indicate that the median observed colours for
the LBG candidates are $J-K=\rm 1.71\pm 0.20$, $i-K=\rm 3.86\pm 0.43$ and
$r-K=\rm 3.43\pm 0.37$. Comparing the $J-K$ colours with the spectra of the
sources, from \S3.3, it is clear that the LBG candidates with the strongest
line emission in the $K$-band also have the reddest continuum colours,
suggesting that the line emission is biasing the colours we measure. With
typical observed-frame equivalent widths of $-$100 to $-$150\,nm, and a
$K$-band filter width of $\sim$350\,nm, the fluxes in $K$ should be corrected
by approximately 0.6--0.7$\times$ (or +0.5 in magnitudes).  No sources are
classed as extremely red objects on the basis of either their $r-K$ or $i-K$
colours, although the reddest source in the optical/near-infrared, SDSS\,J1147,
is also the reddest in the rest-frame UV and is the only candidate LBG in our
sample which does not show Ly$\alpha$ emission. The median $J-K$ colour of our
sample is comparable to that seen for $z\sim\rm 3$ LBGs from \citet{shap01},
$J-K\sim\rm 1.63$, although our candidate LBGs are redder on average in $r-K$
than the standard UV-selected populations at $z\sim\rm 2$ or $z\sim\rm 3$,
$r-K\sim\rm 3.25$ and $r-K\sim\rm 2.85$, respectively \citep{shap01,s04}. This
suggests that the rest-frame UV continua may be significantly redder than
normal LBGs, although their rest-frame optical continua are comparable (before
correcting for emission-line contributions).

Turning to the high-resolution near-infrared imaging (Fig.~1), we find that all
five candidate LBGs are unresolved at the 0.45--0.55-arcsec seeing of our
$K$-band images (as measured from multiple stars in each frame).  Given the
signal to noise of our detections of the LBGs and their measured FWHM relative
to stars in the fields, we can place firm limits of $<0.1$\,arcsec {\sc fwhm}
on the sizes of these sources or on the separation of multiple components if
they are strongly lensed. In the absence of lensing, this angular limit
corresponds to an intrinsic size of $<$1\,kpc for the physical scale of these
sources in their rest-frame $V$-band light.

We identify none of the morphological signatures expected from strong
galaxy--galaxy lensing in our deep $K$-band images: multiple lensed components
or an identifiable foreground lens.  We note that the angular size limit
estimated above, if taken as the Einstein diameter, would correspond to a
velocity dispersion of only 50\,km\,s$^{-1}$ (for a spherical isothermal lens
at $z\sim\rm 0.5$). This velocity dispersion would correspond to a
$\sim$0.1\,L$^\ast$ early-type galaxy with an expected magnitude of $K\sim\rm
19$ at $z\sim\rm 0.5$, detectable in our imaging out to $z\sim\rm 1$
\citep{rus03}. No such nearby lenses are visible in our imaging on the
relevant scales (0.2--2\,arcsec).

Looking at a wider region around the candidate LBGs we see a compact
$K$\,=\,17.9 galaxy (with $J-K\sim\rm 0.8$) lying only 4\,arcsec away from
SDSS\,J1553, but this would not provide a strong amplification of the source.
We also find that SDSS\,J1432 sits in the outskirts of a dense, compact
foreground group, the brightest members of which have $K$\,$\sim$\,16. It is
possible that the LBG candidate is thus magnified by weak lensing from the
foreground structure, although the total magnification is likely to be modest.
The fields surrounding SDSS\,J0243 and SDSS\,J1444 are unremarkable, but we do
identify a group of 5--6 faint, resolved galaxies, $K$\,$\gs$\,19.4, within
15\,arcsec of SDSS\,J1147.  These are possibly members of a cluster, either in
the foreground or (given their faintness) associated with the candidate LBG.
These faint galaxies exhibit a wide range in $J-K$ colours, 0.8--3.0, including
some extremely red objects. In addition, several brighter galaxies,
$K$\,$\gs$\,17.3, lie within 20\,arcsec. We cannot demonstrate a significant
over-density, but they could be more luminous members of the same structure.

Looking at the $(J-K)$--$K$ colour-magnitude plots (Fig.~2) for the five fields
we see that the LBG candidates are rarely the reddest galaxy in the field,
although both SDSS\,J1147 and SDSS\,J1432 have very red, close neighbours
($J-K=\rm 2.2$--3.0).  The most striking feature is the colour-magnitude
sequence seen in the distribution of galaxy colours in the SDSS\,J1432 field.
This has a characteristic colour of $J-K\sim\rm 1.8$ at $K\sim\rm 17.5$,
consistent with that expected for evolved galaxies in a group or poor cluster
at $z\sim\rm 0.7$--0.8 \citep{feul03}. The close similarity of the colours of
the candidate LBG to these galaxies may either be a coincidence or could point
to contamination of our photometric measurement by a superimposed member of
this group (which could thus also gravitationally magnify the background
source).  We believe that the similarities in the colours are merely a
coincidence as it is clear from the spectrum of SDSS\,J1432 (Fig.~3) that there
is a significant contribution to the $K$-band light from the background source.

We conclude that the morphological information for all five of the candidate
LBGs we have imaged at $\ls$0.55-arcsec resolution provides no support for them
being strongly lensed. There is possible evidence for weak lensing for one or
more sources, but this would not significantly affect the apparent magnitudes
of these systems.

\subsection{Spectral properties}

It is immediately apparent from the rest-frame optical spectra presented in
Fig.~3 that all six of the spectroscopically-observed LBG candidates contain
broad-line AGN. All have broad H$\alpha$ emission, with {\sc fwhm} line widths
ranging from 5,000 to $\sim$14,000\,km\,s$^{-1}$ (Table~2) and broad
components visible in H$\beta$ for several systems (despite the typically
poorer signal-to-noise in the $H$-band). SDSS\,J1444 has the weakest emission
and there is no spectral coverage beyond the red extreme of the line due to its
high redshift, but a broad H$\alpha$ line is still apparent. Comparing them to
the two known AGN we observed, the weak-lined AGN SDSS\,J1136 and the
low-ionisation BALQSO SDSS\,J1353, we see that the candidate LBGs exhibit
broader H$\alpha$ emission than either of these two AGN.

So, clear AGN features are visible in the rest-frame optical, whereas the UV
spectra of these galaxies are characterised by a strong continuum but lack the
strong emission lines typical of AGN. Does the AGN contribute significantly to
the UV fluxes of these galaxies?

The redshift measurements available to us tend to follow the same pattern for
all the LBG candidates: the UV-determined values, based predominantly on the
prominent Ly$\alpha$ emission line, are slightly blueward of the [O\,{\sc
iii}]\,500.7nm and H$\alpha$ lines ($\Delta z=$\,0.0051\,$\pm$\,0.0057 and
0.0065$\,\pm\,$0.0037, respectively). The only exception is SDSS\,J1444, where
the Ly$\alpha$ and H$\alpha$ redshifts are identical, but the H$\alpha$
redshift is poorly determined. The H$\alpha$ emission we see originates close
to the AGN, in the broad-line region (BLR), so it is natural to assume that the
UV absorption lines are due to wind-driven material in our line of sight to the
BLR, and that the UV continuum also arises close to the AGN. Indeed, there is a
tight correlation between the FWHM of the H$\alpha$ (Table~2) and the absolute
$i$-band magnitudes from \citet{b04}: the 0.07-dex scatter suggests a close
relationship between the UV continuum emission and the AGN. However, it is not
clear whether this is a direct relationship, or whether it arises merely
because more massive AGN reside in more luminous galaxies. Nevertheless,
assuming that the candidate LBGs have intrinsic power-law continua
characteristic of normal quasars, with $\alpha=-0.44$ \citep{vdb01}, then their
observed rest-frame 200--600-nm spectral slopes ($\alpha = -1.89$ to $-2.54$)
indicate substantial dust extinction, $A_V\sim 1.35$--1.95, for a Calzetti
extinction law \citep{cal00}.

But where are the UV emission lines usually associated with quasar activity? If
they have been quenched by dust surrounding the active nuclei, why can we still
see intense UV continuum emission?  The spectral characteristics of these
galaxies are unusual, but the presence of strong and broad absorption lines in
the UV are similar to those seen in less-reddened examples of the most extreme
BAL quasars found by the SDSS \citep{hall02} and in the Digitized Palomar
Observatory Sky Survey \citep{bru03}. Indeed, very recently \citet{app05} have
published high-resolution \'{e}chelle spectroscopy of SDSS\,J1553 which
provides much higher-quality information about the UV spectral properties of
this galaxy. Based on their analysis of the detailed properties of the
absorption lines, they conclude that SDSS\,J1553 is a low-ionisation BAL quasar
(LoBALQSO), or perhaps an even rarer FeLoBALQSO. The very strong low-ionisation
absorption features found in LoBALQSOs across a wide velocity range can
strongly suppress the emission-line components in these systems, leading to the
absorption-dominanted UV spectra we see.  The UV absorption features of
SDSS\,J1553 are typical of those seen in the other five galaxies and so we
expect that deeper and higher resolution spectroscopy of the complete sample
would likely lead to the same conclusion for the other sources.  Indeed, based
on the existing low-resolution SDSS spectra, \citet{app05} suggest at least
two-thirds of the sample may be LoBALQSOs. The relative weakness and narrowness
of the BAL features, combined with the absence of strong UV emission features,
suggests that the outflows in these galaxies may differ in terms of their
velocity and spatial coverage compared to those seen in typical LoBALQSOs.
Alternatively, these AGN may be similar to SDSS\,J1136 (Fig.~3), which
\citet{hall04} suggest for some unknown reason has weak, broad and highly
blue-shifted emission lines.

Finally, we want to highlight the properties of SDSS\,J1340.  This candidate
LBG was detected at 16\,$\mu$m using {\it Spitzer} by \citet{tep04}.  They
interpret this detection in terms of a massive starburst, even though the
optical/mid-infrared spectral energy distribution (SED) of SDSS\,J1340 in
\citet{tep04} is best fit by the SED for the Seyfert-1, NGC\,5548. Our
non-detection of the source in the submm suggests that the 16\,$\mu$m detection
most likely arises from high-temperature AGN-heated dust, rather than a
bolometrically luminous starburst. Further support for the presence of a
bolometrically luminous AGN in this system comes from the detection of a strong
and broad H$\alpha$ line in our near-infrared spectrum (Fig.~3).

\section{Conclusions}

We present multi-wavelength observations of a sample of six candidate LBGs at
$z=\rm 2.5$--2.8 identified from the SDSS DR1 QSO Catalog by \citet{b04}.  We
suggest that these sources could be either: 1) intrinsically luminous,
UV-bright starbursts; 2) strongly-lensed examples of typical-luminosity LBGs;
or 3) a class of quasars with extremely weak UV emission lines.

We do not detect any of the four candidate LBGs observed in the submm, placing
a strong constraint on the submm emission from the ensemble. This suggests that
the sources are unlikely to be strongly-lensed examples of more typical LBGs,
or intrinsically-luminous LBGs, unless the far-infrared emission from such UV
starbursts declines precipitously at high luminosities. Two further pieces of
evidence weigh against the lensing hypothesis: first, the UV spectral
properties of the candidate LBGs do not match those of typical luminosity LBGs;
second, using high-resolution near-infrared imaging of five of the candidates
we find no morphological evidence of strong lensing. Taking these results
together, we conclude that the sources in our sample are unlikely to be either
intrinsically-luminous LBGs or rare, strongly-lensed examples of more normal
LBGs. This suggests that they are most likely to be unusual AGN.

Our near-infrared spectroscopy confirms this suggestion, identifying very broad
lines in the rest-frame optical spectra of all six galaxies in the sample.  We
therefore conclude that the six apparently extremely luminous LBGs identified
by \citet{b04} are likely to be LoBALQSOs whose unusually weak UV emission
lines may either be an intrinsic property of these AGN \citep{hall04}
or result from a complex distribution of absorption in the outflow close to
the AGN.

\section*{Acknowledgments} 
We thank Pat Osmer and David Weinberg for their work on the SDSS LBG survey.
We also thank Alastair Edge for useful conversations, and the referee for
suggestions that improved the paper markedly. We acknowledge service
observations from the JCMT. IRS acknowledges support from the Royal
Society. MB is supported by a Graduate Fellowship from the National Science
Foundation. AWB acknowledges support from NSF grant AST-0205937, the Research
Corporation and the Alfred P.\ Sloan Foundation.

\end{document}